\documentclass[a4paper]{jpconf}
\usepackage{color}
\usepackage{graphicx}
\begin{document}
\title{Recent Heavy Flavor Results Utilizing the FVTX in PHENIX at RHIC}

\author{Xuan Li$^{[1]}$ for the PHENIX Collaboration}

\address{1) P-25, MS H846, Physics Divison, LANL, Los Alamos, NM, 87545, USA}

\ead{xuanlipx@rcf.rhic.bnl.gov}

\begin{abstract}
Heavy flavor and quarkona production are important hard probes to test the Quantum Chromodynamics (QCD) and measure the properties of the Quark Gluon Plasma (QGP) created in high energy heavy ion collisions. 
The new PHENIX mid- and forward-rapidity silicon vertex detectors provide precise Distance of Closest Approach (DCA) determination, allowing charm and bottom yields identification, and improve di-muon mass resolution for charmonia studies.
The relative $\psi(2S)$ over $J/\psi$(1S) yields measured in $p$+$p$, $p$+Al and $p$+Au data at $\sqrt{s}=$200 GeV help us to understand final state effects on the charmonia production. Direct access to B-meson production has been enabled via measurements of non-prompt $J/\psi \to \mu^{+} + \mu^{-}$ in $p$+$p$ collisions at $\sqrt{s} = 510$ GeV and Cu+Au collisions at $\sqrt{s} = 200$ GeV. 
The fraction of B-meson decays in J/$\psi$ yields in the $p$+$p$ measurement agrees with higher energy results from Tevatron and LHC. This fraction is enhanced in Cu+Au collisions suggesting the B-meson yield is not modified by the medium whereas J/$\psi$ is strongly suppressed.
\end{abstract}

\section{Introduction}
Because of their high mass ($m_{\rm{c,b}} \gg \lambda_{\rm{QCD}}$), heavy quarks are produced in the early stage of high energy collisions. Since the heavy flavor quarks do not vanish or change into other flavors during the hard scattering and fragmentation processes, they can be treated as hard probes to study the full evolution of the medium created by relativistic heavy ion collisions. Heavy flavor particles that traverse the medium in heavy ion collisions experience both cold and hot nuclear matter effects. To understand the hot nuclear matter effects in heavy ion collisions, it is necessary to study the same measurements in $p$+$p$ and $p$+A collisions as well. 

The Forward Silicon Vertex Tracker (FVTX)  \cite{fvtx_nim} installed in PHENIX since 2012 at RHIC dramatically improves the tracking quality in the rapidity range of $1.2<|y|<2.2$, with full azimuthal angle coverage. Better mass resolution of di-muon pairs reconstructed through the FVTX now allows separation of $\psi(2S)$ from $J/\psi$(1S) in the FVTX rapidity range. 
The mid-rapidity $\psi(2S)$ measured in $\sqrt{s}$ = 200 GeV d+Au collisions is more suppressed compared with the $J/\psi$ measurement in the same data set \cite{psi_dau}. 
This result suggests that the final state effect is not negligible, as both $J/\psi$ and $\psi(2S)$ are expected to undergo the same initial state effects. Extension to forward/backward rapidity measurements will further establish how final state effects impact the charmonia with different bound state energies. 

The PHENIX mid-rapidity charm and bottom separated single electron $R_{\rm{AA}}$ measured in 200 GeV Au+Au collisions suggests that the bottom production is less suppressed than charm production in the electron $p_{T}<4$ GeV/c region \cite{phenix_cb_sep}. This measurement is consistent with models which expect mass dependency of quark energy loss \cite{Eloss} and flavor dependent dissociation and formation time \cite{I_Vitev}. D/B mesons measured in the forward/backward rapidity may probe a different QGP evolution stage in heavy ion collisions from the mid-rapidity region. Decayed particles with a non-zero displaced vertex can be identified with the help of the FVTX. Direct access to B-mesons in the forward/backward rapidity is first studied at PHENIX in both 510 GeV $p$+$p$ collisions and 200 GeV Cu+Au collisions through the B $\rightarrow$ $J/\psi$ $\rightarrow$ $\mu^{+}$$\mu^{-}$ decay channel. 

\section{The double ratio of $\psi(2S)$ over $J/\psi$ measured in $p$+$p$, $p$+Al and $p$+Au collisions at $\sqrt{s}$ = 200 GeV}
Yields of $J/\psi$ and $\psi(2S)$ have been measured in $p$+$p$, $p$+Al and $p$+Au collisions at $\sqrt{s}$ = 200 GeV from the RHIC 2015 run. The double ratio of $\psi(2S)$ to $J/\psi$, which is defined as the ratio of $\psi(2S)$ yields over $J/\psi$ yields measured in $p$+Al and $p$+Au collisions over the same ratio measured in $p$+$p$ collisions, is used to characterize relative nuclear modifications between $\psi(2S)$ and $J/\psi$.

\begin{figure}[!ht]
\begin{center}
	\includegraphics[width=0.48\linewidth]{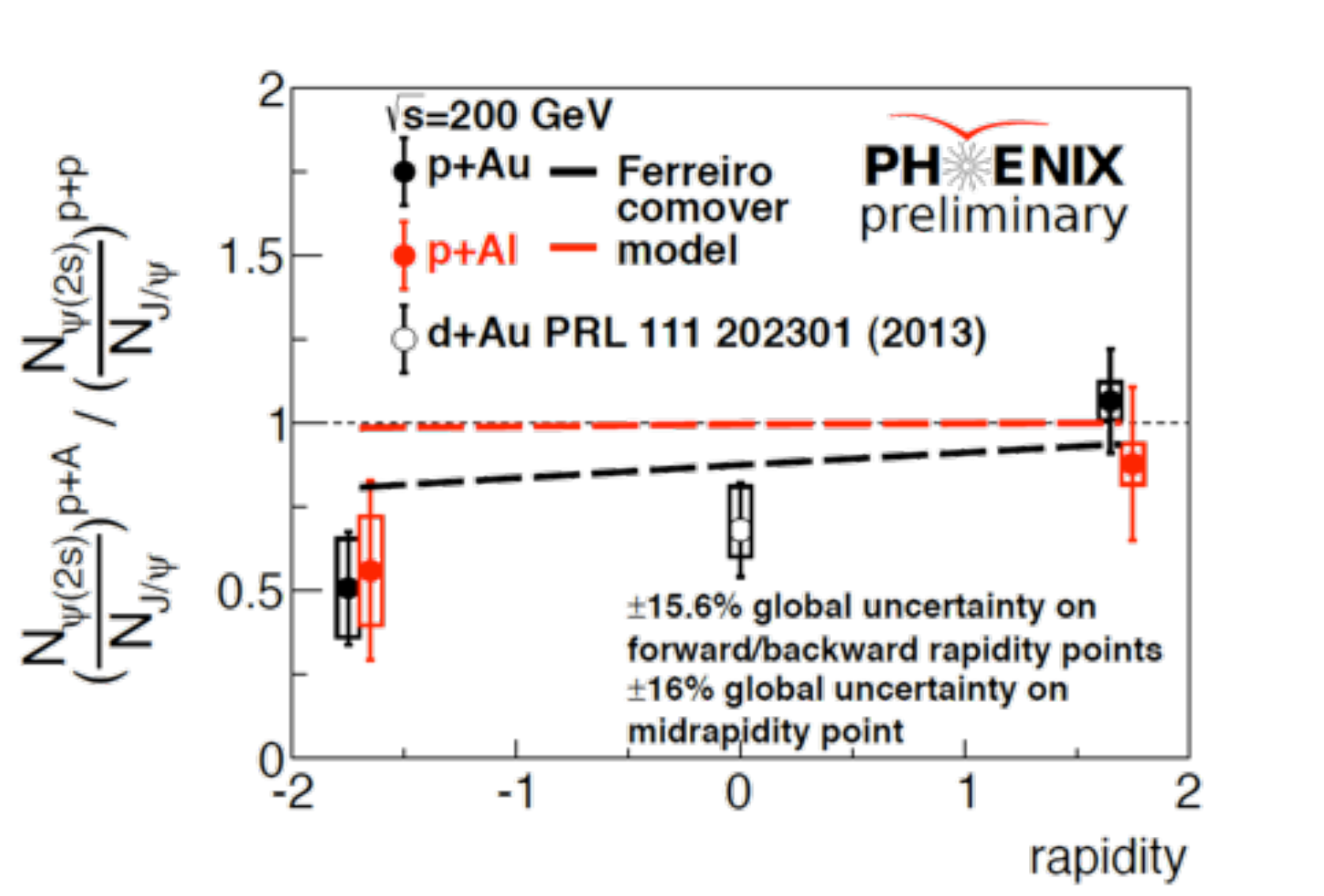} 
	\includegraphics[width=0.48\linewidth]{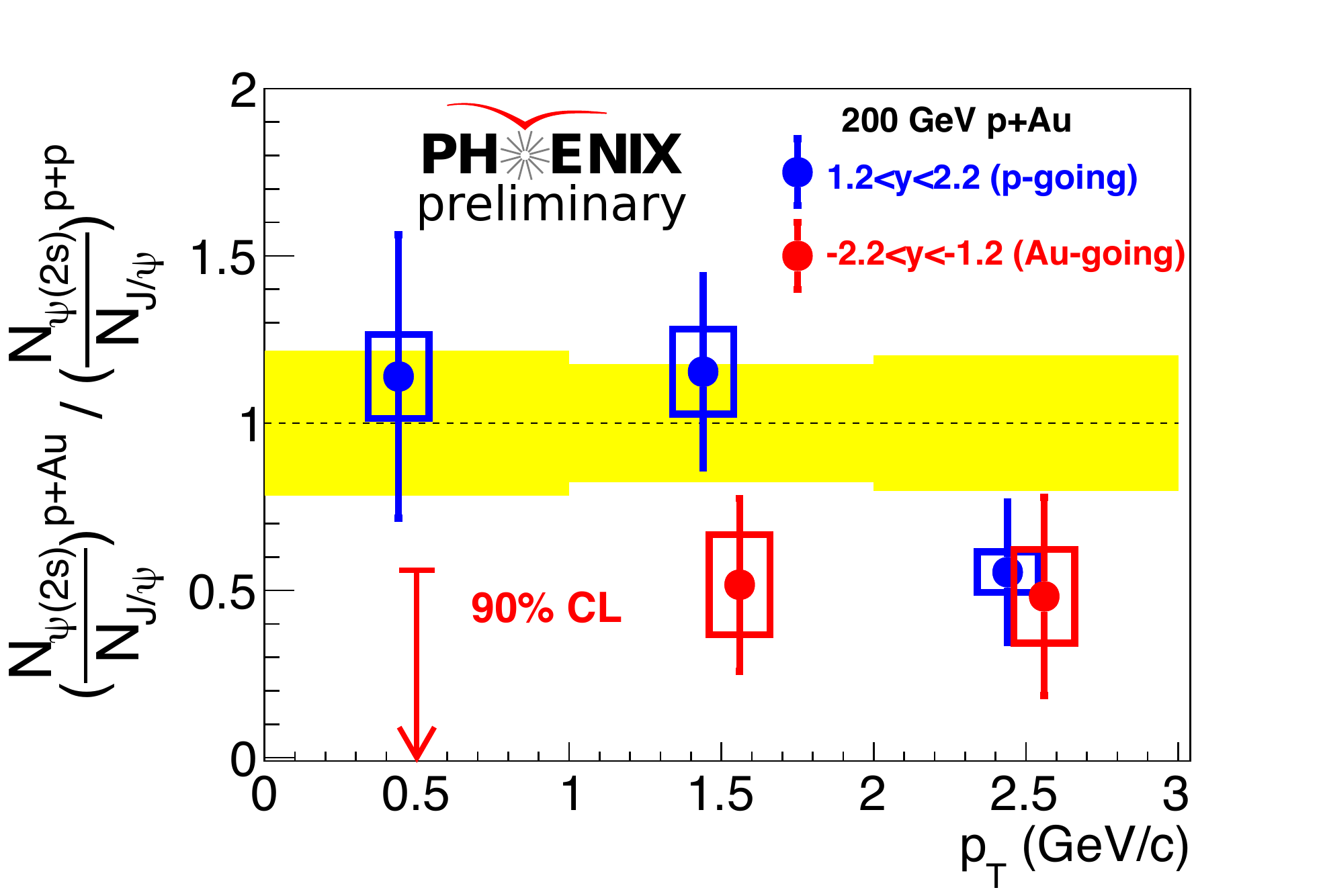}
	\caption{\label{fig:jpsi_psi_double_ratio} Left: The rapidity dependent double ratio of $\psi(2S)$ to $J/\psi$ measured between $p$+Au, $p$+Al and $p$+$p$ collisions at 200 GeV in comparison with the co-mover model calculation \cite{E_Ferreiro}. Right: The $p_{T}$ dependent double ratio of $\psi(2S)$ to $J/\psi$ measured in $p$+Au collisions at 200 GeV.}
\end{center}
\end{figure}

The forward rapidity (p-going) $\psi(2S)$ to $J/\psi$ double ratios measured in both $p$+Au and $p$+Al collisions are near unity. This means $\psi(2S)$ to $J/\psi$ have similar nuclear modifications in the proton going direction, where particle densities are smaller. In the backward rapidity, significant decreasing of the $\psi(2S)$ to $J/\psi$ double ratios is measured in $p$+Au and $p$+Al collisions indicating the loosely bounded $\psi(2S)$ is suppressed more than $J/\psi$ in the Au/Al going direction, where particle densities are larger. Comparison with co-mover model calculations \cite{E_Ferreiro} are shown in the left panel of Figure \ref{fig:jpsi_psi_double_ratio}. Reasonable agreement between data and the co-mover model on the rapidity dependence means the charmonia break up with co-mover mechanism might be the dominant contributor to the $\psi(2S)$ and $J/\psi$ differential suppression. The $p_{T}$ dependent $\psi(2S)$ to $J/\psi$ double ratio is studied in $p$+Au collisions as shown in the right panel of Figure \ref{fig:jpsi_psi_double_ratio}. 
Absence of low $p_{T}$ $\psi(2S)$ in the backward rapidity may be caused by the longer time interaction with soft co-movers.

\section{The fraction of $J/\psi$ from B-meson decay in $\sqrt{s}$ = 510 GeV $p$+$p$ collisions and $\sqrt{s}$ = 200 GeV Cu+Au collisions}
The fraction of $J/\psi$ from B-meson decay is first studied in the RHIC 2012 510 GeV $p$+$p$ collisions and 200 GeV Cu+Au collisions. Identification of $J/\psi$ from B-meson decay at forward/backward rapidity ($1,2<|y|<2.2$) relies on the Distance of Closet Approach (DCA) measurement with the FVTX and VTX detectors at PHENIX. Prompt particles (such as prompt J/$\psi$) show up as symmetric DCA distributions, while the DCA distribution of long lived decayed particles (such as B-meson decays) produces a long tail for negative DCA and an asymmetric distribution.

The analysis use muons from unlike-sign di-muons in the J/$\psi$ mass region. 
The DCA shapes of muons which decay from prompt $J/\psi$ and B-meson decayed $J/\psi$ are determined from a PYTHIA+GEANT+reconstruction simulation and/or samples of identified prompt hadrons. 
Combinatorial, detector mis-matching and heavy flavor continuum backgrounds are determined separately using real data. A fit of the DCA distribution can simultaneously determine the prompt $J/\psi$ and non-prompt $J/\psi$ from B-meson decays.

\begin{figure}[!ht]
\begin{center}

	\includegraphics[width=0.4\linewidth]{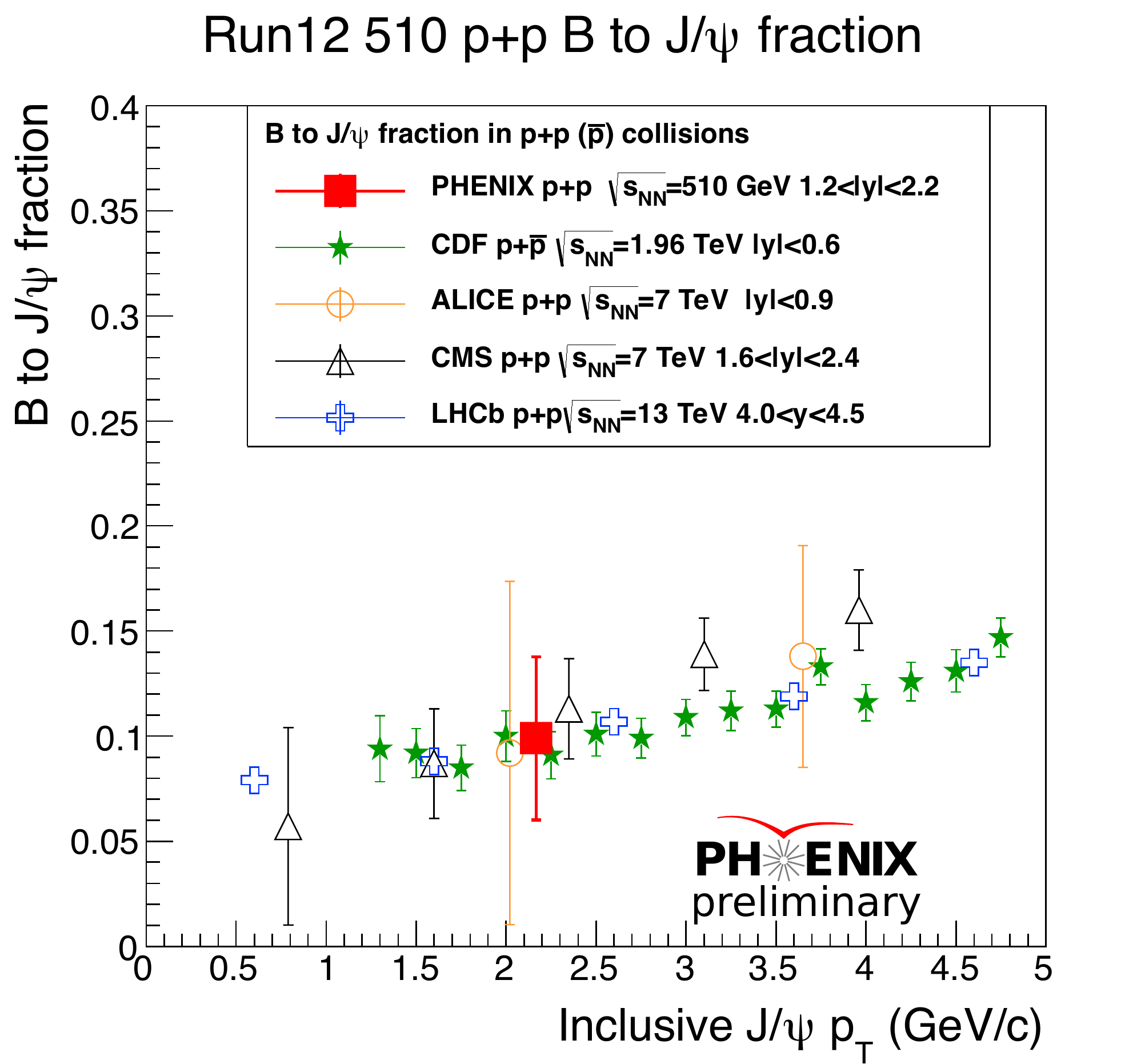}
	\includegraphics[width=0.4\linewidth]{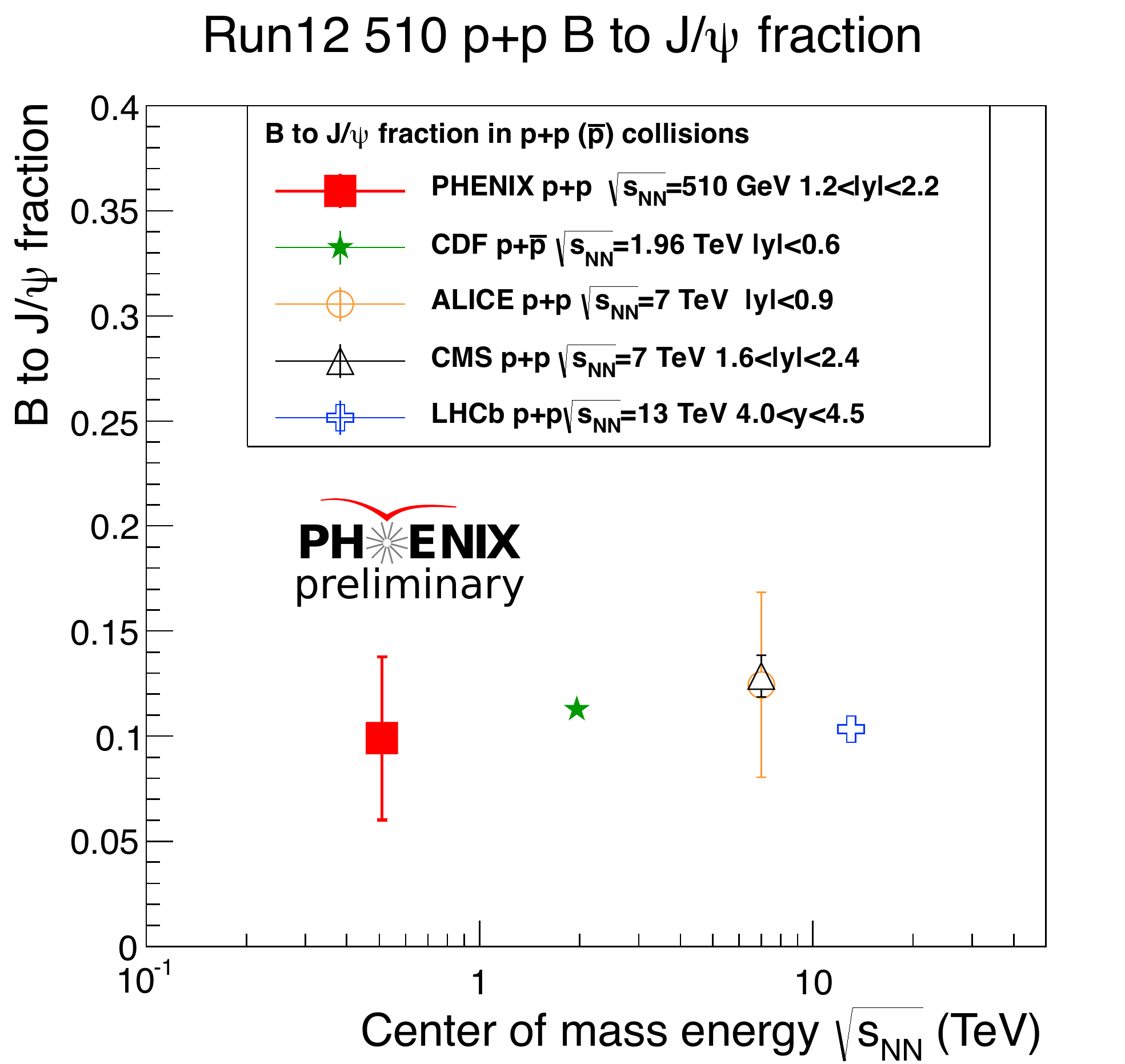}
	\caption{\label{fig:btojpsi_510pp} The fraction of $J/\psi$ from B-meson decays measured in 510 GeV $p$+$p$ collisions at PHENIX and comparison with the global data from CDF \cite{Acosta:2004yw}, ALICE \cite{Abelev:2012gx}, CMS \cite{Khachatryan:2010yr} and LHCb \cite{Aaij:2015rla} experiments versus inclusive $J/\psi$ $p_{T}$ (left) and versus center of mass energy (right). }
	\end{center}
\end{figure}

\begin{figure}[!ht]
\begin{center}
	\includegraphics[width=0.49\linewidth]{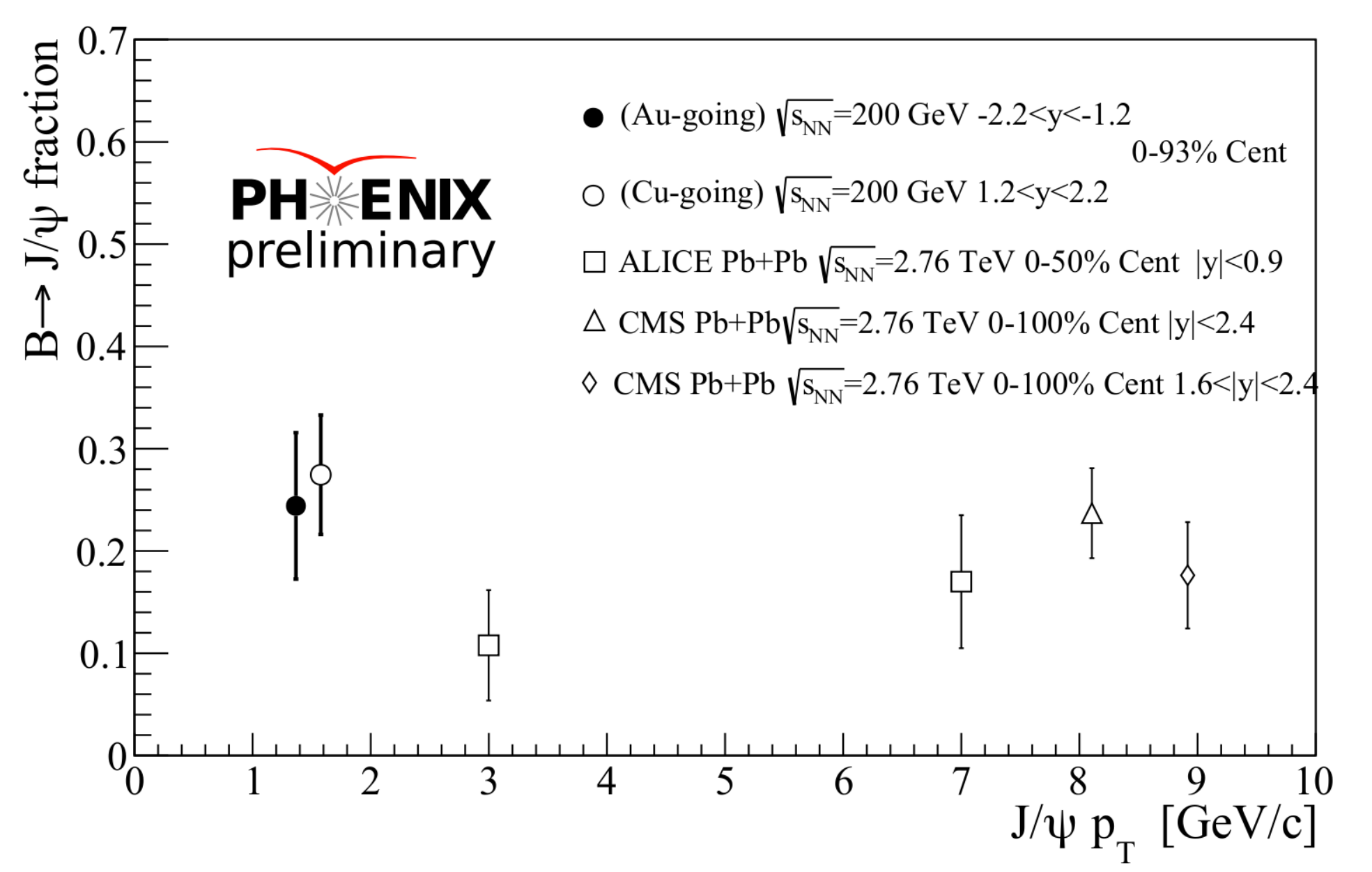}
	\includegraphics[width=0.42\linewidth]{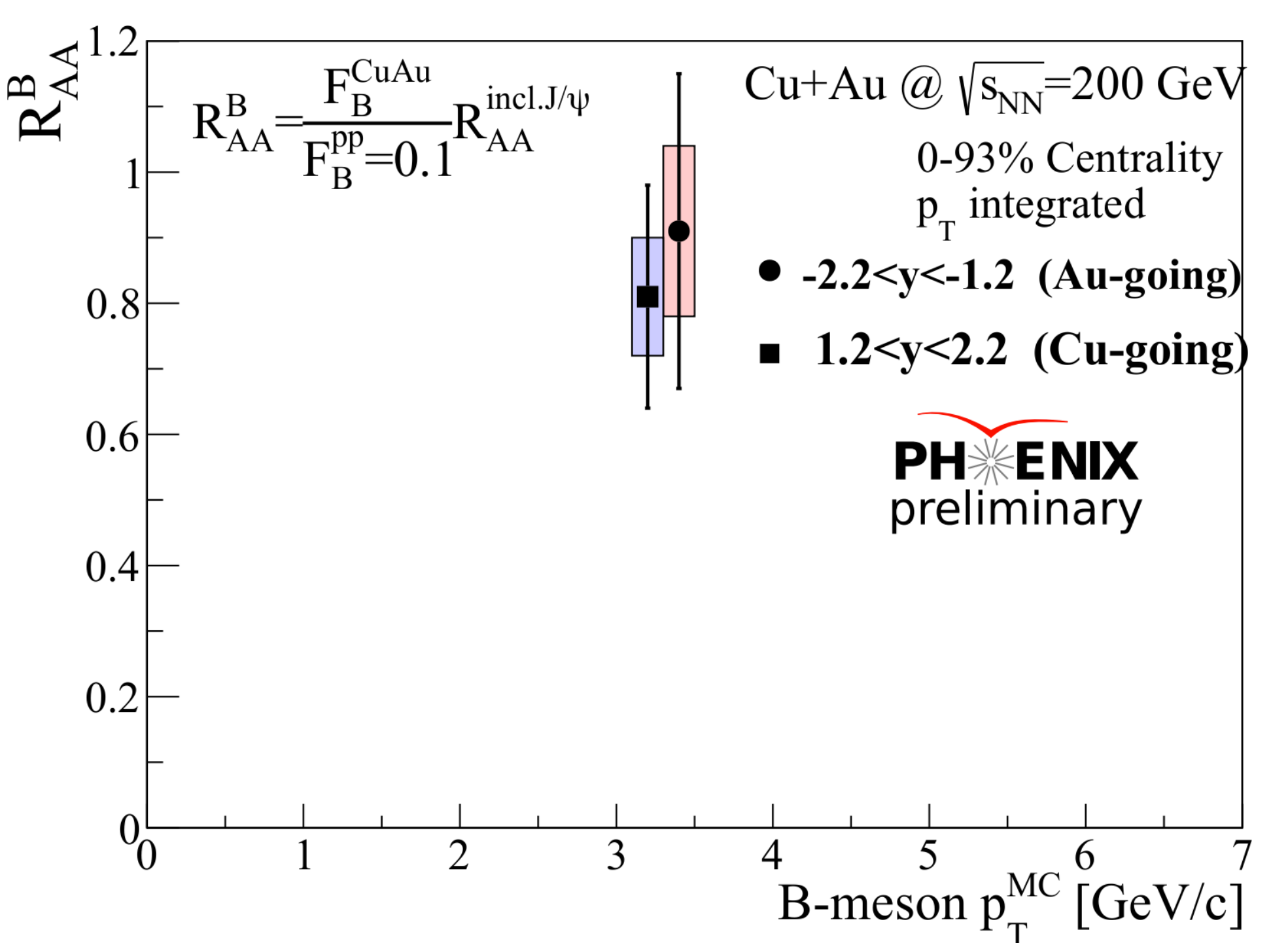}
	\caption{\label{fig:btojpsi_200CuAu} Left: The fraction of $J/\psi$ from B-meson decay versus inclusive $J/\psi$ $p_{T}$ measured in 200 GeV Cu+Au collisions in comparison with the results measured in 2.76 TeV Pb+Pb collisions from ALICE \cite{alice_pbpb}, and CMS \cite{cms_pbpb} experiments. }
\end{center}
\end{figure}

The detector acceptance*efficiency corrected fraction of $J/\psi$ from B-meson decay measured in 510 GeV $p$+$p$ collisions is shown in Figure \ref{fig:btojpsi_510pp}. This result is compared with other measurements within the same inclusive $J/\psi$ $p_{T}$ region ($p_{T}<5$ GeV/c) at different rapidities and center of mass energies. The PHENIX preliminary result follows the $p_{T}$ dependence of the global data. The B-meson decayed $J/\psi$ fraction does not have a strong energy dependence in $p$+$p$ or $p$+$\bar{p}$ collisions for the $J/\psi$ $p_{T}<5$ GeV/c region. 

Figure \ref{fig:btojpsi_200CuAu} shows the PHENIX preliminary result of $J/\psi$ from B-meson decay fraction in Cu+Au collisions. The fractions in the Cu-going direction and Au-going direction are consistent with each other within large uncertainties.
The $J/\psi$ from B-meson decay fraction measured by PHENIX in Cu+Au collisions at $\sqrt{s_{NN}}=$200 GeV is larger than results measured at the LHC in 2.76 TeV Pb+Pb collisions, indicating that B-mesons have weaker nuclear modification compared to the one from J/$\psi$ at RHIC energy.
To better understand this, the nuclear modification factor $R_{\rm{CuAu}}$ of $J/\psi$ from B-mesons decay is determined. As the 200 GeV p+p analysis is ongoing, we assume the fraction is 0.1 in 200 GeV p+p collisions according to the scaling behavior shown in the right panel of Figure \ref{fig:btojpsi_510pp}. 
We calculate the $R_{\rm{CuAu}}$ of $J/\psi$ from B-mesons decay (shown in the right panel of Figure \ref{fig:btojpsi_200CuAu}) by applying the measured B-meson decay $J/\psi$ fraction in Cu+Au collisions and the assumed value in $p$+$p$ collisions to the measured inclusive $J/\psi$ $R_{\rm{CuAu}}$. 
The result is consistent with no nuclear modification of momentum and centrality integrated B-meson yields. The FVTX detector provides almost uniform acceptance for a broad momentum range of B-mesons starting from zero $p_T$. This measurement is dominated by low $p_T$ B-mesons.

\section{Summary and Outlook}
Differential suppression of the $\psi(2S)$ to $J/\psi$ double ratio in 200 GeV $p$+Al and $p$+Au data shows strong evidence of final state effects in the backward rapidities. The co-mover model provides a reasonable description. First measurements of $J/\psi$ from B-meson decay in forward/backward rapidity at low $p_{T}$ are achieved for both 510 GeV p+p and 200 GeV Cu+Au collisions at PHENIX. No significant center of mass energy dependence for low $p_{T}$ $J/\psi$ from B-meson decay in p+p collisions is observed for energies of 510 GeV and above. No significant nuclear modification of integrated $p_{T}$ and centrality B-mesons is observed in 200 GeV Cu+Au data, which is consistent with a hard scattering process and no additional production in the medium. 

Large data sets in various types of heavy ion collisions collected by PHENIX provide opportunities to further study the heavy flavor productions in forward/backward rapidities. The $\psi(2S)$ to $J/\psi$ double ratio in 200 GeV Au+Au collisions will access the hot nuclear matter effects such as color screening in QGP. The method to study the DCA of muons in the B-meson decays to J/$\psi$ fraction analysis will be applied to other data sets and to the study of B- and D-meson semi-leptonic decays to muons in the forward/backward rapidities. Studies in $p$+Au and Au+Au collisions will improve the understanding of the cold and hot nuclear matter effects.

\section{References}
\bibliographystyle{iopart-num}
\bibliography{xuanli}
\end{document}